\documentclass[preprint]{revtex4}

\begin{document}

\title{On the vacuum entropy and the cosmological constant\footnote{This essay received an
``honorable mention" in the 2003 Essay Competition of the Gravity
Research Foundation.}}

\author{Saulo Carneiro}

\affiliation{Instituto de F\'{\i}sica, Universidade Federal da
Bahia, 40210-340, Salvador, BA, Brazil}

\begin{abstract}
It is generally accepted that the entropy of an asymptotically de
Sitter universe is bounded by the area, in Planck units, of the de
Sitter horizon. Based on an analysis of the entropy associated to
the vacuum quantum fluctuations, we suggest that the existence of
such a holographic bound constitutes a possible explanation for
the observed value of the cosmological constant, theoretically
justifying a relation proposed $35$ years ago by Zel'dovich.
\end{abstract}

\maketitle

As extensively discussed in the literature \cite{9}, there is a
fundamental problem related to the existence of a positive
cosmological constant. If $\Lambda$ originates from vacuum energy,
its expected value, on the basis of quantum field calculations
with a cutoff given by the Planck energy, has the order of
$l_{Planck}^{-2} \approx 10^{70}$m$^{-2}$. This is $122$ orders of
magnitude greater than the observed value $\Lambda \approx
10^{-52}$m$^{-2}$ \cite{8}. Even considering smaller cutoffs, as
the energy scales of electroweak or QCD phase transitions, the
expected value is still over 40 orders of magnitude too high. This
huge discrepancy is known as the cosmological constant problem.

We have shown elsewhere \cite{GS,preprint} that this problem can
be related to other open issues in cosmology, as the large numbers
coincidence and the cosmic coincidence problem, with the help of
the holographic principle \cite{11}. In this essay, we will try to
show that the application of this principle to an asymptotically
de Sitter universe leads to a value for $\Lambda$ in accordance
with observation.

In a simplified form, the holographic principle can be described
as the extension, to any gravitating system, of the
Bekenstein-Hawking formula for black-hole entropy. More precisely,
it establishes that the number of degrees of freedom of the system
is not bounded by its volume in Planck units, as expected from
quantum theories of space-time, but by the area, in Planck units,
of its delimiting surface.

The use of this principle to the universe as a whole depends on
the definition of such a surface at a cosmic scale. The existence
of a positive cosmological constant naturally introduces a
characteristic surface of radius $\Lambda^{-1/2}$ \cite{18}.
Another possibility is to use the Hubble horizon, with radius
$H^{-1}$, for it defines the scale of causal connections for any
observer \cite{17}. It is clear that in a homogeneous and
isotropic, infinite universe, filled with dust and a positive
cosmological constant, this last version implies the former, for
the Hubble radius tends asymptotically to $\sqrt{3/\Lambda}$.
Therefore, in the context of an asymptotically de Sitter universe,
the holographic bound establishes that the maximum number of
available degrees of freedom is given by
\begin{equation}
\label{Nmax} N_{max} \approx \Lambda^{-1}.
\end{equation}

How much entropy has our universe? We know that the number of
baryons is of the order of $10^{80}$, and that the cosmic
background radiation contains about $10^{8}$ photons per baryon.
It is reasonable to believe that dark matter contributes with a
similar figure. But the major contribution to the entropy of
matter seems to come from massive black-holes present in galactic
nuclei, which represent an entropy of the order of $10^{101}$
\cite{Penrose}.

But what about the entropy associated to the vacuum fluctuations?
To put this question in a proper way and to clarify its role in
solving the cosmological constant problem, let us inquire more
carefully on the origin and meaning of this problem.

The difficulties appear when we calculate the vacuum energy
density with the help of quantum field theories in flat
space-time. The vacuum energy comes from two kinds of
contributions. The first one is the energy associated to the
vacuum expectation value of self-interacting fields, as the Higgs
field, the quark and gluon condensates of QCD, or any other field
associated to vacuum phase transitions. The second kind comes from
the zero-point fluctuations of the fields, which lead to infinite
results. To regulate them, it is used to impose some energy
cutoffs, but, as said before, this leads to results still many
orders of magnitude too high compared to observation.

Nevertheless, as discussed by some authors (see for example
\cite{Ralf}), any contribution, infinite or not, to the vacuum
energy density predicted by quantum field theories in flat
space-time must be exactly canceled by a bare cosmological
constant in the Einstein equations, because in the flat space-time
the right-hand side of those equations is identically zero.
Therefore, to properly pose the problem, we have to calculate the
vacuum energy density in a curved background. As before, we find a
divergent result as well. But now a physically meaningful
(renormalized) value for $\Lambda$ should be obtained by
subtracting the Minkowskian result. Since the space-time of our
universe is not strongly curved at cosmic scale, we expect to
obtain a small value for $\Lambda$, in accordance with
observations.

The situation is analog to what occurs in the Casimir effect.
There, the zero-point fluctuations of the electromagnetic field
give rise to an infinite contribution to the vacuum energy
density, inside and outside the region between the Casimir plates.
But what is physically meaningful, leading to observable effects,
is the difference between the values in the two regions, which is
shown to be finite. In our case, the role of Casimir plates is
played by gravity.

Calculating the vacuum energy density in a curved background is a
difficult task. An example of a rough estimation in the line of
the above reasoning was recently given by Sch\"utzhold
\cite{Ralf}. He estimates the contribution for $\Lambda$ from the
chiral anomaly of QCD in a curved, expanding space-time, obtaining
$\Lambda \approx H \Lambda_{QCD}^3$, where $\Lambda_{QCD}$ is the
energy scale of the chiral phase transition. In the limiting case
of a de Sitter universe, $H \approx \sqrt{\Lambda}$, and his
result leads to
\begin{equation}
\label{Lambda} \Lambda \approx m^6
\end{equation}
(where we have made $m = \Lambda_{QCD}$).

This expression was derived $35$ years ago by Zel'dovich, from
empirical arguments \cite{Zel'dovich}. Although it is sensible to
the parameter $m$, (\ref{Lambda}) leads to the correct order of
magnitude: using $\Lambda_{QCD} \approx 150$ MeV, we obtain
$\Lambda \approx 10^{-51}$ m$^{-2}$, in good agreement with
observation, considering that it was not taken into account
numerical factors.

An alternative to circumvent the difficulties involved in quantum
field calculations in a curved background is to use a
thermodynamic approach, which does not depend on the details of
the field dynamics. The idea is to obtain a superior limit for the
vacuum entropy, instead of its energy density, and to compare the
result with the holographic bound (\ref{Nmax}). The reader may
argue that it is not trivial to define the number of virtual
particles in curved backgrounds. Let us remind, however, that our
universe has a quasi-flat space-time. Therefore, the estimation
given below can be considered a good approximation.

It is clear that non-trivial vacuum configurations of classical
fields (as the vacuum expectation value of the Higgs field or the
QCD condensates) do not contribute to the vacuum entropy. In what
concerns the zero-point fluctuations, they have, properly
speaking, an infinite entropy density, because (if we do not
impose any energy cutoff) the number of modes is infinite. But if
we regulate their energy, by introducing an ultraviolet cutoff
$m$, we also regulate their entropy. A simple estimation of the
resulting entropy bound can be derived as follows.

Limiting the energy-momentum space associated to the zero-point
fluctuations leads to the quantization of their configuration
space, with a minimum size given by $l \approx m^{-1}$. This
results in a superior bound to the number of available degrees of
freedom in a given volume, say, the volume inside the Hubble
horizon. The maximum number $N$ of observable degrees of freedom
will be of the order of $V/l^3$, where $V$ is the Hubble volume.
That is,
\begin{equation}
\label{N} N \approx \left(\frac{m}{H}\right)^3.
\end{equation}

Now, if we take for $H$ the de Sitter asymptotic value $H \approx
\sqrt{\Lambda}$, we can identify (\ref{N}) with the holographic
bound (\ref{Nmax}). It is easy to verify that this leads to
Zel'dovich's relation (\ref{Lambda}).

But why $m$ coincides to the energy scale of the QCD phase
transition? The common belief is that a natural cutoff should be
given by the Planck energy, for at the Planck scale the classical
picture of space-time breaks down. Nevertheless, it is not
difficult to see that, equating (\ref{Nmax}) to (\ref{N}), with
$H\approx\sqrt{\Lambda}$ and $m=m_{Planck}$, one obtains a Hubble
radius of the order of $l_{Planck}$, which is not consistent with
our universe.

One can also argue that the zero-point fluctuations of other
fields than quarks and gluons contribute to the entropy as well.
It is then intriguing that just $\Lambda_{QCD}$ enters in
Zel'dovich's relation. Note, however, that in a curved space-time
the different sectors of the standard model of particles
interactions are coupled by gravity. On the other hand, the de
Sitter universe is a stationary space-time, and, therefore, all
the (interacting) vacuum fields should tend to a state of
thermodynamic equilibrium, at the temperature of the last vacuum
phase transition. But the last of such transitions was the chiral
transition of QCD, at a temperature given by $\Lambda_{QCD}$.

Finally, let us note that the vacuum entropy already dominates the
universe entropy. Indeed, taking for $H$ the value observed
nowadays, $H \approx 65$ km/(sMpc) \cite{8}, we obtain from
(\ref{N}) (with $m \approx \Lambda_{QCD}$) $N \approx 10^{122}$, a
figure that predominates over the matter entropy referred above,
of order $10^{101}$.

${}$

I am thankful to F.S. Navarra, M.R. Robilotta, A. Saa, A.E.
Santana, and R. Sch\"utzhold, for useful discussions. I am also
grateful to Prof. P.F. Gonz\'alez-D\'{\i}az, G.A. Mena Marug\'an
and L. Garay, from CSIC (Madrid), where I have started to work
with this subject.

\end{document}